\documentstyle[psfig,prl,twocolumn,aps,floats]{revtex}

\begin{document}

\draft
\widetext

\title{Metallic bonding due to correlations: A quantum chemical ab-initio calculation
of the cohesive energy of mercury} 

\author{Beate Paulus$^1$ and Krzysztof Rosciszewski$^{1,2}$}

\address{$^1$ Max-Planck-Institut  f\"ur Physik komplexer Systeme,
N\"othnitzer Stra\ss e 38, D-01187 Dresden, Germany}

\address{ $^{2}$ Institute of Physics, Jagellonian University, Reymonta 4,
Pl 30-059 Krakow, Poland}

\maketitle

\begin{abstract}
Solid mercury in the rhombohedral structure is unbound within the self-consistent field (Hartree-Fock)
approximation. The metallic binding is entirely due to electronic correlations.
We determine the
cohesive energy of solid mercury within an ab-initio many-body
expansion for the correlation part. 
Electronic correlations in the $5d$ shell contribute about half to the cohesive energy.
Relativistic effects are found to be very important.
Very good agreement with the experimental value is obtained.
\end{abstract}
Pacs-No.: 71.15.-m, 71.20.-b, 96.30.Dz 
\vspace*{1cm}

\narrowtext

\begin{figure}
\begin{center}
\psfig{figure=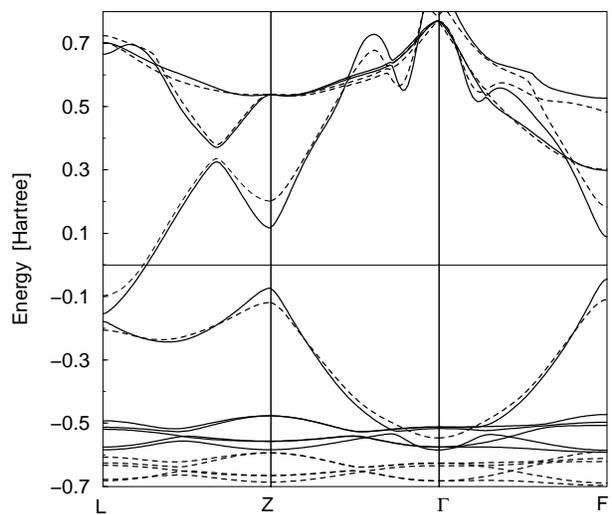,angle=-90,width=7.5cm}
\end{center}
\caption{\label{band}
The Hartree-Fock band structure of mercury at the experimental lattice constant of the rhombohedral structure.
The Fermi energy is set to zero seperately for the relativistic (full line) and for the non-relativistic calculations (dashed line).}
\end{figure}

Mercury is a very special metal: It is a liquid at room temperature and it freezes at $T=233$K 
with a rhombohedral lattice ($a_0=3.005$\AA, $\alpha=
70.53^{\rm o}$\cite{crc97}).
The mercury dimer is a van der Waals bound molecule with a binding energy of about 0.05 eV\cite{zehnacker87},
more than one magnitude
lower than the bulk cohesive energy of 0.79 eV per atom\cite{crc97}.
Increasing the size of mercury clusters, the character of binding changes from van der Waals over covalent (around 20 atoms) to
metallic (more than 100 atoms)\cite{haberland90}.\\
The origin of the cohesion in the solid mercury metal is the topic of this letter. A mean-field Hartree-Fock treatment yields no
binding: In contrast to the textbook knowledge of the metallic bond, the cohesion of mercury is entirely due to correlations. 
No surprise, the cohesive energy and the lattice constants of bulk
mercury have been a long outstanding problem in computational physics\cite{singh94a,singh94b,moyano02}. 
Questions of particular interest are
the influence of the $d$-shell correlation on the binding and the importance of relativistic effects.
For the dimer, the $d$-shell correlation contributes with more than a factor of 2 to the dissociation energy\cite{dolg96,schwerdtfeger01}. 
Wavefunction-based methods with very large basis sets achieve good
agreement with experiment. 
For the metallic solid, one could expect DFT calculations
to give reasonable results, but different functionals yield quite different values of the cohesive energy\cite{moyano02}.
As dimer or trimer can be handled with wavefunction-based methods, it seems
natural to employ a many-body expansion for the cohesive properties of larger clusters or
the solid. However, a straight-forward many-body expansion with  2-body and 3-body forces taken from
free clusters works properly for rare gas crystals\cite{roscis99} but fails for metallic mercury\cite{moyano02}.\\
We propose a combined approach: The self-consistent field treatment is performed at the
Hartree-Fock (HF) level in the periodic 3-dimensional infinite solid\cite{crystal98} and the correlation energy is
calculated within a many-body incremental scheme\cite{stoll92b}.
The latter embeds larger and larger fragments of the solid with shells of the same atoms and the same
geometrical parameters, but describes them with a smaller basis. 
Thus the contribution of the metallic binding is calculated at a self-consistent field level in the infinite solid,
while the correlation part, which is fairly local, is treated in finite embedded clusters 
with wavefunction based correlation
methods.\\
The Hartree-Fock calculation is performed with the program package Crystal98\cite{crystal98}. To deal with the 
scalar-relativistic effects, we apply a small-core scalar-relativistic pseudopotential\cite{andrae90}
where the $5s^2p^6d^{10}6s^2$ electrons are treated explicitly. As basis set for Crystal98
we modify the valence-double-zeta (vdz, i.e. providing 2 atomic orbitals for each occupied crystal orbital) 
correlation-consistent basis set
of Peterson\cite{peterson03}
resulting in $(6s6p6d)/[4s4p3d]$. We frequently compare with results of a calculation with
a non-relativistic small-core pseudopotential\cite{andrae90}.\\
A first important observation is that mercury is not bound at the HF level (see Table \ref{coh}), 
both in the relativistic and the non-relativistic case. 
We checked that the system is unbound within the HF approximation for a wide range of lattice constants. Thus,
the binding is entirely due to correlations. The cohesive energy at the HF level is less repulsive in the non-relativistic treatment. 
This is due to the smaller atomic $sp$ splitting in the latter case. The 
$p$ orbitals are too low in energy and their contribution to binding is overestimated.\\ 
The relativistic HF band structure (see Fig.\ref{band})
and the density of states near the Fermi level are quite similar
to the one obtained with LDA\cite{singh94b,deng98}. Between the L and Z points a conduction band 
crosses the Fermi level, responsible for the metallic behaviour.
The logarithmic singularity which occurs with the HF approximation at
the Fermi level is not discernible for the $k$-mesh used. 
The $5d$ bands are quite flat and placed well below the Fermi level. 
They overlap strongly with the $6s$ band at the
$\Gamma$ point, indicating
the importance of the core-valence correlations. The strong $sp$ mixing can be seen in the 
$\ell$-decomposed density of states. Near the Fermi level, both have nearly the same magnitude.
A Mulliken population analysis for the solid yields an electron transfer of 0.73 $e$ from the
$s$ shell to the $p$ shell,  compared to 1.03 $e$ in the nonrelativistic case.
The non-relativistic HF band structure (dashed lines in Fig.\ref{band}) shows a smaller
$sp$-splitting at $\Gamma$ than the relativistic calculation. In a two-band model, one expects therefore the
hybridization gaps, e.g., at the L or F point, to be larger and the $p$
character of the valence band to be stronger, as it is seen in the bulk calculation (Fig.\ref{band}).\\
We now proceed to the correlation treatment with the incremental scheme\cite{stoll92b}. The correlation energy of the solid
is expanded in 1-body increments $\epsilon_i$, 2-body increments $\Delta \epsilon_{ij}$ and so on:
\begin{equation}
\label{ecorr}
  E_{\rm corr} = \sum_i             \epsilon_i +
                 \sum_{i<j  }\Delta \epsilon_{ij} +
                 \sum_{i<j<k}\Delta \epsilon_{ijk} + \ldots
  \quad .
\end{equation}
The sums include groups of occupied localised orbitals.
Note that we expand only the correlation part, in contrast to 
Moyano et al.\cite{moyano02}, whose many-body expansion involves the total energy. A second difference 
is that we perform our calculations in embedded clusters to mimic
the confinement of the electrons in the solid; for comparison see
\onlinecite{roscis04}.\\
The incremental scheme obtains the energy increments from finite clusters calculations.
For solid mercury the clusters were generated as follows: The rhombohedral structure 
can be viewed as a central atom surrounded by atom shells of various size.
There first shell contains 12 atoms,
6 of them at distance $a_0$ and 6 at 1.15$a_0$. 
We select the atoms which we want to correlate and surround each of them 
with atoms of the first shell. If the atoms to be correlated are far away, we select for the embedding
also all atoms found within a cylinder of radius 1.15$a_0$ around the connection line.
The embedding atoms are described with a large-core 
pseudopotential\cite{kuechle91} and only with a $s$ basis set $(4s)/[2s]$\cite{kuechle91}.
This small basis set prevents the electrons to move onto the surface of the cluster,
but still mimics the Pauli repulsion of the neighbourhood sufficiently well.
Performing a HF calculation of the cluster (the basis set for the small-core pseudopotential is [3s1p1d], i.e, without 
unoccupied $p$ basis functions) and localising
the occupied orbitals\cite{foster60}, yields well
localised orbitals on the individual atoms, which can be separated in embedding orbitals
and orbitals to be correlated.
With this procedure, we can generate local van der Waals like orbitals of the solid structure.\\
The delocalization of true metallic orbitals is treated via the incremental scheme as follows:
Any orbital group $i$ contains now all localised orbitals of atom $i$.
In the first step we enlarge the basis set of the atom to be correlated to a reasonable quality,
recalculate the integrals and reoptimize only the orbitals of the atom  to be correlated in a
HF calculation while the orbitals of the embedding are kept frozen. This provides us with fairly local
orbitals on the atom to be correlated. On top of this HF calculation, we perform 
a Coupled-Cluster calculation with Single and
Double excitations and perturbative treatment of the Triples (CCSD(T))\cite{hampel92}. All calculation
of the finite clusters are performed with the program package MOLPRO2002\cite{molpro2002}.
The correlation energy provides us the 1-body increment $\epsilon_i$.\\
The cohesive part of the 1-body increment
\begin{equation}  
\Delta\epsilon_i^{\rm coh}=\epsilon_i-\epsilon^{\rm corr}_{\rm atom}.
\end{equation}
is defined as the change of the correlation energy of the atom due 
to surrounding it with other atoms.
The correlation energy of the free atom $\epsilon^{\rm corr}_{\rm atom}$ 
is calculated with the same basis set as used in the finite cluster (basis functions on the surrounding, too).
We list the cohesive part of the 1-body increment for different orbitals correlated and
different basis sets applied in the first part of Table \ref{basis}. Basis A utilizes on Peterson's vdz basis $(7s7p6d1f)/[7s6p5d1f]$.
Basis B utilizes on Peterson's vtz exponents and corresponds to $(10s9p7p2f1g)/[8s7p6d2f1g]$
and Basis C on the vqz exponents, i.e. $(12s12p9d3f2g1h)/[9s8p7d3f2g1h]$.
All values of $\Delta\epsilon_i^{\rm coh}$ are repulsive. The atom in the metallic solid has a smaller
correlation energy than the free atom, because the cage effect increases the level spacing and therefore decreases
the correlation energy. For a reasonable basis set, e.g.
Basis B or Basis C, where the Hilbert space is large enough to polarise the closed $5d^{10}$ 
and the $5s^2p^6$ shell, the correlation of the $5d^{10}$ shell is the same in the free atom
and in the atom in the solid, but the correlation of the underlying $5s^2p^6$ shell yields 
in the solid a repulsive contribution of about 0.8 mH.\\
Next, we consider the contributions from the 2-body increments. All pairs of atoms
up to 3.0$a_0$ are embedded separately and, again,  
HF calculations are performed to generate
the localised orbitals. The embedding part is kept frozen, while the orbitals of the atoms to be correlated 
are reoptimized with a better basis set and then correlated at the CCSD(T) level. The correlation of the $5d^{10}6s^2$ shell yields
the 2-body correlation energy $\epsilon_{ij}$. Only the non-additive part 
\begin{equation}
\label{twobody}
\Delta\epsilon_{ij}=\epsilon_{ij}-\epsilon_i - \epsilon_j .
\end{equation}
is required for the correlation energy of the solid
(Eq. \ref{ecorr}). 
To  minimize the error due to the embedding, the 1-body increment
which is subtracted in Eq.~(\ref{twobody}) has been determined in the same cluster with the second atom 
described with the small basis set. 
We have performed the calculation of the 2-body increments with Basis B up to a distance of
3.0$a_0$, where the increment is less than 10 $\mu$H. We fit the van der Waals formula to the region  
from 1.5$a_0$ to 2.8$a_0$. The value $C_6$=297 atomic units is comparable with data from the literature
\cite{schwerdtfeger01,hartke02,kunz96}.
Now, we can estimate the cohesive energy contributions from far-away pairs. If
we include explicitly  increments from all pairs smaller than 2.52$a_0$ and estimate the rest (see Table \ref{coh}), the latter
contribute only about 2\% to the 2-body cohesive energy.\\
Table \ref{basis} lists the nearest-neighbour 2-body increment for different basis sets and different 
orbitals to be correlated. Pure valence correlations contribute about 40\% (Basis B) to the two-body increment, 
the $5d^{10}$ shell 
correlations are splitted in the 
$5d^{10}$-valence (38\%) and $5d^{10}$-$5d^{10}$ (22\%) correlation (Basis B).
The correlation of the $5s^2p^6$ shell increases the magnitude of the 
2-body increment by 6\%. The increase of the 2-body increment applying Basis C instead
of Basis B is about 7\%. 
But probably still the basis set limit is not reached. The dissociation energy
of the free Hg dimer obtained with an augmented vqz basis set, which is comparable to our Basis C for the solid,
is still 13\% below the estimated basis set limit\cite{peterson03}.\\
The next step of the many-body expansion are the 3-body increments. It is well known that 
3- and higher-body contributions are important for the binding in mercury\cite{moyano02}. We calculate all
3-body increments where two distances are $\leq 1.65a_0$
and the third $\leq 3.30 a_0$. The 3-body increment
\begin{equation}
\Delta\epsilon_{ijk}=\epsilon_{ijk}-(\Delta\epsilon_{ij}+\Delta\epsilon_{ik}+\Delta\epsilon_{jk})\\
-(\epsilon_{i}+\epsilon_{j}+\epsilon_{k}) 
\end{equation}
is only the non-additive part due to the simultaneous correlation ($5d^{10}6s^2$ shell) of three atoms in the solid, and therefore
expected to be small in comparison with the 2-body increment.
The corresponding 2- and 1-body increments are calculated in the same embedding
as the 3-body correlation energy $\epsilon_{ijk}$. We apply Basis B for the 3-body increments. 
Summing all 3-body increments where two distances are $\leq 1.15a_0$ yields
an attractive contribution to the cohesive energy of 5.1 mH, whereas the rest yields only 2.4 mH.
If we follow Axilrod and Teller\cite{axilrod43} and fit a dipole-dipole-dipole interaction 
to the 3-body increments where no distance is shorter than 1.5$a_0$ and estimate
the far away triangles with this fitting formula, the sum over the far-away triangles contributes less than 0.01 mH to the cohesion.
This is due to the rapid decay of $r^{-9}$ and  a substantial cancellation of contributions from triangles with different angles.
The basis set dependence of the 3-body increments was tested for the largest individual 3-body increment.
Applying Basis C instead of Basis B yields an increase of 20\%.\\
The 4-body contributions are calculated according to the scheme described for the 3-body increments.
Here Basis A is applied. For one geometry Basis B is used, the increase of the increment is about 20\%. 
Five connected 4-body clusters have been selected.
The largest contribution comes from the linear one. Like for the linear 3-body increment, the 
presence of a highly polarisable central atom yields a
significant increase of the correlation energy. But even for the large linear increments, the magnitude of the 
4-body increment is by a factor of 4 smaller than the corresponding 3-body increment. In addition, most 
4-body increments are repulsive, so we can expect an alternating sign series for the 
correlation energy of the solid. \\
Besides the contributions discussed up to now there are 
two small contributions to the cohesive energy resulting from the spin-orbit coupling and the
zero-point energy. For the former, published values\cite{dolg96,schwerdtfeger01} for the dissociation energy of the Hg dimer 
show only a small influence.
The spin-orbit coupling of the $5d$ shell should be not so much different in the solid and the free atom,
therefore the net effect on the cohesive energy should be small.\\
The latter one is due to zero-point phonon degrees of freedom.
An estimation of the zero-point energy can be done with the Debye model 
$E_{\rm ZPE}=\frac{9}{8}k_{\rm B}\Theta_{\rm D}$\cite{farid91}, with the Debye temperature $\Theta_{\rm D}$=71.9K
for mercury\cite{kittel}.\\ 
We have summarized our results in Table \ref{coh}. The HF part of the cohesive energy is
repulsive, the 1-body correlation contribution to the cohesive energy is repulsive, too. 
The main contribution to the binding comes from the 2-body increments, with about half of this
originating from the core-valence correlation of the $d$ shell. Without correlating the $d$ shell
solid mercury would not be bound. The non-relativistic 2-body increments
yield a significant over-binding. The 3-body contributions of the valence shell are attractive, whereas
the 3-body contributions of the $d$ shell are repulsive, reducing the 3-body contributions
to about 10\% of the 2-body contributions. Furthermore, the non-relativistic 3-body terms are
strongly over-binding. This is due to the enlarged valence correlations and to the fact
that the non-relativistic $d$ shell is not interacting with the $s$-band at the $\Gamma$
point (see Fig.\ref{band}). Both lead to a too small repulsive part due to the core-valence correlation.
The spin-orbit contribution and the zero-point energy can be neglected. 
If we extrapolate the 2-body-contributions to Basis C and to the inclusion of $5s^2p^6$ correlation
we reach 74\% of the experimental cohesive energy. If we further estimate the basis set error 
to the basis set limit from the dimer data, very good agreement with experiment (99\%) is obtained.\\
In summary, we have presented a method, which allows us to determine the cohesive energy of solid mercury within the same accuracy
with the 
experimental value as it was achieved for the much simpler rare gas crystals\cite{roscis99}. 
It can be easily systematically improved by expanding to higher orders and by applying better
basis sets when the computing capabilities improve. Of course, the combination of extended HF calculations and
the incremental scheme for the correlations can be applied to other metallic systems, too. 
But the specific nature of the mercury-mercury bond constitutes arguably the most interesting case.\\[1cm] 
{\bf Acknowledgement:} The authors would like to thank Hermann Stoll (Stuttgart, Germany), Peter Fulde 
(Dresden, Germany) and Erich Runge (Dresden, Germany)
for many valuable discussions and Peter Schwerdtfeger (Auckland, New Zealand), who first drew their attention to the 
mercury puzzle.

\begin{table}
\begin{tabular}{|ll|rrr|}
&Basis&    $6s^2$   &   $5d^{10}6s^2$   &  $5s^25p^65d^{10}6s^2$    \\ \hline
$\Delta\epsilon_1^{\rm coh}$& A& +0.004287  &   +0.004740 & +0.005390 \\
&B&  +0.004230  &   +0.004260 & +0.005089 \\
&C&+0.004168  &   +0.004210 & +0.0049801\\
\hline 
$\Delta\epsilon_{12}$&A&  -0.003468  &  -0.009648  &   -0.010291\\
&B& -0.004107   &  -0.009862 &   -0.010487\\
&C& -0.004278   &  -0.010550 &   -0.011268
\end{tabular}
\caption{\label{basis} The correlation contribution to the cohesive energy of the 
1-body increment $\Delta\epsilon_1^{\rm coh}$ and the nearest-neighbour 2-body increment
$\Delta\epsilon_{12}$ (both in Hartree) are listed above for different 
basis sets (detailed description in the text) and
different orbitals to be correlated.}
\end{table}

\begin{table}
\begin{tabular}{|lrrr|}
& $6s^2$& $5d^{10}6s^2$&$5d^{10}6s^2$\\
&rel.&rel.& non-rel.\\
\hline
Hartree-Fock& +36.2&+36.2&+17.4\\
\hline 
$\Delta\epsilon_i^{\rm coh}$&+4.2&+4.3&+3.9\\
2-body up to 2.52$a_0$&-25.0&-50.0&-62.0\\
far away 2-body&& -1.1&\\
ext. to Basis C and $5s^2p^6$ corr.&& -6.6&\\
3-body in the first shell&-10.3& -5.1&-15.7\\
3-body in the second shell&& -2.4&\\
est. for far away 3-body&& $\le$ -0.1&\\
4-body (selected clusters)&& +3.0&\\
spin-orbit&&$\le$ -0.1&\\
zero-point energy&& +0.3&\\
\hline
calc. cohesive energy&& -21.4&\\
est. to basis limit (from dimer)&&-7.5&\\
exp. cohesive energy && -29.0& 
\end{tabular}
\caption{\label{coh} The different contributions to the cohesive energy of mercury in mHartree
per unit cell.} 
\end{table}

\end{document}